# From webtables to datatables

Mária Csernoch
University of Debrecen, Hungary, 4028 Debrecen Kassai út 26.
csernoch.maria@inf.unideb.hu

## ABSTRACT

*Webtables – tables and table-like structures on webpages – are excellent sources for teaching spreadsheeting, in commercial and professional organisations by utilizing and developing knowledge-transfer items, presenting and handling various real-world problems and solutions, discussing and debugging, and in general, developing and utilizing computational thinking skills. In the present paper the conversion process of one of the LOL Boards (League of Legends, Riot Games Inc. 2019) is detailed. After presenting the algorithm of the conversion, two solutions are offered – one in a word processor, the other purely in a spreadsheet application – leaving space for discussions, inventing other solutions and combining them.*

## 1. INTRODUCTION

Teaching spreadsheeting from the programming point of view (Sestoft, 2011) is an approach which evokes various knowledge-transfer items which have traditionally not been focused on in end-user computing. The programming aspect allows us to pay attention to data analysis, data types, n-dimensional vectors, n×m dimensional matrixes– without the burden of explicit declaration –, file and data conversions, handling n-ary and multilevel functions, algorithms, and coding in general concept-based (Pólya, 1957; Csernoch, 2017) high-mathability (Baranyi & Gilányi, 2013; Biró & Csernoch, 2015a, 2015b) computer problem-solving approaches.

To make problem- and programming-oriented spreadsheeting classes efficient and effective, teachers must present data which students are interested in, which suit their level of background knowledge and their age, and which they are eager to use for information retrieval. In short, teachers must make students motivated. However, most of the tutorials available are not like this. They are long (hundreds of pages), cover non-spreadsheet knowledge items (how to handle ribbons and toolbars, how to setup the spreadsheet environment, how to format and colour data cells and tables, how to handle operating system features, how to type data, etc.), lack any real world problem-solving, and are claimed to be suitable for everyone (Katz, 2010; Walkenbach, 2010; ECDL, 2016; ICAEW, 2016; Job Test Prep, 2016; Microsoft, 2016; Test ECDL, 2019).

In teaching-learning practice, instead of relying on ready-made tutorials, teachers must select contents which best suit their students' interests. One option is to work with the students' data collected in other classes in school or at work. However, these data are not necessarily available and/or do not match the students' background knowledge. One solution is to turn to the internet and download and modify contents which match all the requirements.

The present paper details how webtables are selected and converted into datatables and how the conversion process can be used for developing students' and participants' computational thinking skills by applying knowledge-transfer items. In this context webtables are structures in which data are presented in tables or table-like forms on webpages, while datatables are $n{\times}m$ matrixes with $n$ data records and $m$ data-fields, and usually with an extra row for fieldnames.

## 2. WEBTABLE→DATATABLE CONVERSION: THE BACKGROUND

Webtable→datatable conversion requires various ICT skills which in traditional teaching approaches are presented as software-specific elements of knowledge (discussed in Csernoch, 2017; Csernoch & Biró, 2018).This approach is clearly detectable in most teaching materials (Katz, 2010; Walkenbach, 2010; ECDL, 2016; JobTestPrep, 2016; Microsoft, 2016; Test ECDL, 2019), in widely accepted exams (ECDL/ICDL, SAM), and even in the Spreadsheet Competency Framework (ICAEW, 2016).

The aim of the present paper is to demonstrate how different applications and background knowledge deriving from various sources would lead to (1) long lasting knowledge, (2) deep approach problem-solving with slow thinking, (3) reliable fast thinking problem-solving based on schemata, and (4) a realisation of algorithms and their importance in computer problem-solving, regardless of interfaces.

### 2.1. Selecting webtables

In the teaching-learning process, especially at the beginning, the selection of webtables and datatables and how they are presented is primarily the teacher's responsibility (Kirschner et al., 2006). The content and the form in which tables are introduced in class greatly affects the outcome. In no particular order of importance, the following points of view must be taken into consideration when authentic data sources are presented in class.

– Using authentic tables in class would save students/participants (students for short) from the labour involved in typing (typing is not a spreadsheet competence; furthermore, it is boring, time consuming, unreliable, small-sized tables discredit the use of spreadsheets, etc.). In our approach, typing is not banished, but mostly applied in creating formulas (coding in spreadsheet interfaces), instead of typing data.

– Content plays a crucial role in motivating students. Tables which students are interested in and which contain data from which they are willing to gain information would best serve our purposes. Another option would be to handle various data collected in non-ICT classes. However, this second option, at the time of writing of the present paper, is hardly used, due to the low level of spreadsheet knowledge, both on the part of students and non-computer teachers. Consequently, computer teachers are responsible for providing data sources which are compatible with other subjects and sciences and match the students' interest, age, occupation, etc.

– A third aspect of selecting tables is how these data sources are presented in class. The form in which the data is available for students is primarily affected by the aims of the class. From pure web pages to normalized tables, and any state in between– all are acceptable.

– The fourth aspect is the issue of timing, i.e. when these tables should be introduced in spreadsheet classes. The answer is simple: they should be introduced right at the very beginning, during the very first class. It has been found that the first class plays an important role in persuading students that they really need spreadsheeting. If we are able to make them believe that they need this knowledge, that they can handle large worksheets from the very beginning, and that they can even form questions and provide answers with formulas, we can win.

## 2.2. The challenges of conversions

In our experience, the webtable→datatable conversion (WeT→DaT) started as a 'must' for presenting authentic data in classes. However, in the course of events, we have found that the conversion processes might serve as a webpage semantic validator (Csernoch & Dani, 2017). Building the algorithms for the conversion and using traditional methods for handling different file formats (4.1.1 and 4.1.2 in Table 1, 4.3.2 in Table 2, 5.1.1, 5.1.2, 5.1.4, 5.1.5 in Table 5) might reveal the discrepancies of webtables: redundancy, the amount of data stored in one data-field, the structures used to build the tables or table-like formations, how consequently data are recorded and stored, the characters used in the tables (especially spaces, paragraph and line breaks, decimal and thousand separator characters), character coding, etc. We must note here that importing data from web pages with the user-friendly Data→From Web command works properly only on well-structured and well-designed webtables, which is not often the case. The method presented in this paper would serve as an alternative solution for handling poorly designed webtables also. It might look "old-fashioned", but its advantage to the ready-made solution is that it is program, version, and data-structure independent, furthermore highly supports knowledge-transfer activities, built upon the content of long-term memory (Sweller et al., 2011).

This deeply felt need to present interesting and processable tables in spreadsheet classes became a passion. Encountering a new webtable is extremely exciting. In hyper attention mode (Csernoch & Dani, 2017) it is impossible to tell what challenges the selected webtable holds. Is it convertible or not? If it is convertible, would we be able to do it or not? What new tricks and algorithms are required to solve problems specific to the actual table? The webtable→datatable conversion process has turned out to be a great challenge, and we are addicted to it.

## 2.3. Selection of the application

As mentioned above, it is always the teacher's responsibility to decide in which form tables are presented in class. This statement also implies that all the tables must be tried and converted beforehand. Furthermore, all conversions must be remade shortly before the class. This extra caution must be taken due to the ever-changing structure and content of webpages. Without checking the webtables in detail before the class takes place, they might surprise us and lower the efficacy of the teaching-learning process (e.g. the IMDb Top 250 table has gone through several changes over the years and various versions are available: IMDB, 2019a, 2019b).

One of the most fundamental knowledge-transfer items in end-user computing is that there is life beyond opening a datafile by double clicking, based on the datafile-application assignment. Relying on this knowledge, webpages can be opened both in spreadsheet and word processor applications. However, opening a webpage in a spreadsheet application requires more attention due to the automated, and frequently undesired, datatype recognition, which in European languages (i.e. not English) can cause serious conversion problems. In Figure 1–Figure 3 (Social Blade LLC, 2019) the comma as the thousand separator character is the source of the conversion problem (Figure 3). A test was carried out by adding 1 to the converted values (Figure 3). In the example, the Excel-converted values are stored in Columns C–E, while the results of the addition are shown in Columns F–H ({=C2:C251+1}, {=D2:D251+1}, and {=E2:E251+1}, respectively) (Figure 2–Figure 3).

[Table image showing webtable with columns: Rank, Grade, Username, Uploads, Subs, Video Views]

*Figure 1. Top 250 youtubers in the United Kingdom – original webtable.*

[Excel screenshot showing columns B-H with Rank, Uploads, Subs, Views, Uploads_test, Subs_test, Views_test]

*Figure 2. Top 250 youtubers in the United Kingdom – opened and tested in English Excel.*

[Excel screenshot showing the same data with #ÉRTÉK! errors in Subs_test and Views_test columns]

*Figure 3. Top 250 youtubers in the United Kingdom – opened and tested in a European (Hungarian) Excel.*

In Figure 4–Figure 6 (Fatty Weight Loss, 2019) the decimal-character-problem is presented. The original webpage uses a dot as the decimal character, which is a reasonable decision, given the language of the page. However, the webpage, with the left alignment format, presents these data as string, so the conversion process would tell how these data are stored.

Opening the webpage in an English version of Excel is a convenient start in the conversion process and provides correct data in a fast way (Figure 5). All the text contents are recognized with a string, while all the numbers with number data types. However, opening the webpage in a non-English Excel converts decimal real numbers to dates (Figure 6; Cells D2, F2, F3, and F314) or strings (Figure 6, Columns D and E, except Cell D2 and Cells F4, F312, F313, and F315).

| Description | Serving | Calories (kcal) | Protein (g) | Fat (g) | Carbohydrates (g) | Cholesterol (mg) |
|---|---|---|---|---|---|---|
| Abiyuch, raw | .5 cup | 69 | 1.5 | 0.1 | 17.6 | |
| Acerola juice, raw | 1 cup | 23 | 0.4 | 0.3 | 4.8 | 0 |
| Acerola, (west indian cherry), raw | 1 cup | 32 | 0.4 | 0.3 | 7.69 | 0 |
| Apple juice, canned or bottled, unsweetened, with added ascorbic acid | 1 cup | 47 | 0.06 | 0.11 | 11.68 | 0 |
| Apple juice, canned or bottled, unsweetened, without added ascorbic acid | 1 cup | 47 | 0.06 | 0.11 | 11.68 | 0 |
| Apple juice, frozen concentrate, unsweetened, diluted with 3 volume water without added ascorbic acid | 1 cup | 47 | 0.14 | 0.1 | 11.54 | 0 |

*Figure 4. Fatty Weight Loss. Food Calorie Tables. Fruits and Fruit Juices. – original webtable.*

When the real numbers are converted into dates, there is no way back, the original data is not available anymore, due to the conversion process operating in the background, in which dates are converted into whole numbers. Data being converted into a string datatype is a more favourable solution.

| | A | C | D | E | F | G |
|---|---|---|---|---|---|---|
| 1 | Description | Calories (kcal) | Protein (g) | Fat (g) | Carbohydrates (g) | Cholesterol (mg) |
| 2 | Abiyuch, raw | 69 | 1.5 | 0.1 | 17.6 | |
| 3 | Acerola juice, raw | 23 | 0.4 | 0.3 | 4.8 | 0 |
| 4 | Acerola, (west indian cherry), raw | 32 | 0.4 | 0.3 | 7.69 | 0 |
| 312 | USDA Commodity pears, canned, ligh | 62 | 0.28 | 0.15 | 16.21 | |
| 313 | USDA Commodity, mixed fruit (peach | 57 | 0.46 | 0.1 | 14.65 | 0 |
| 314 | USDA Commodity, mixed fruit (peach | 55 | 0.41 | 0.08 | 14.3 | 0 |
| 315 | Watermelon, raw | 30 | 0.61 | 0.15 | 7.55 | 0 |

*Figure 5. Fatty Weight Loss. Food Calorie Tables. Fruits and Fruit Juices. – opened in English Excel.*

| | A | C | D | E | F | G |
|---|---|---|---|---|---|---|
| 1 | Description | Calories (kcal) | Protein (g) | Fat (g) | Carbohydrates (g) | Cholesterol (mg) |
| 2 | Abiyuch, raw | 69 | 05.jan | 0.1 | 17.jún | |
| 3 | Acerola juice, raw | 23 | 0.4 | 0.3 | 08.ápr | |
| 4 | Acerola, (west indian cherry), raw | 32 | 0.4 | 0.3 | 7.69 | 0 |
| 312 | USDA Commodity pears, canned, | 62 | 0.28 | 0.15 | 16.21 | |
| 313 | USDA Commodity, mixed fruit (pea | 57 | 0.46 | 0.1 | 14.65 | 0 |
| 314 | USDA Commodity, mixed fruit (pea | 55 | 0.41 | 0.08 | 14.márc | 0 |
| 315 | Watermelon, raw | 30 | 0.61 | 0.15 | 7.55 | 0 |

*Figure 6. Fatty Weight Loss. Food Calorie Tables. Fruits and Fruit Juices. – opened in a European (Hungarian) Excel.*

### 3. WEBTABLE→DATATABLE CONVERSION: LOL BOARD

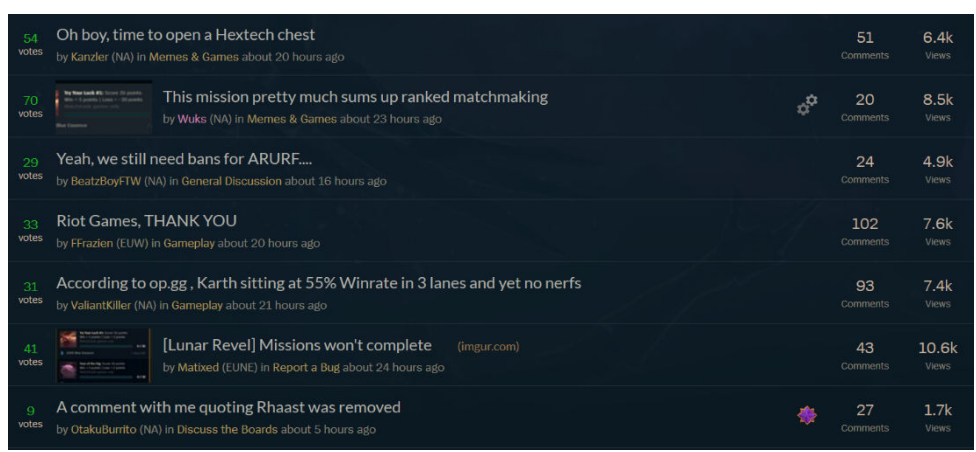

*Figure 7. The opening page of one of the LOL boards.*

In the following, the major issues involved in a conversion process are presented. For this demonstration, we selected a webtable which middle- and high-school students are extremely interested in, namely the LOL boards(League of Legends, Riot Games Inc., 2019) (a 14-old student suggested using this webtable) (Figure 7). The table can also serve didactic purposes, since it (1) contains several data fields with different datatypes, (2) redundancy can be detected, (3) pictures in different positions and with various meanings are displayed, and (4) minor lemmatization problems must also be handled.

### 3.1. At first sight– hyper attention mode

At this stage of the conversion a data-scanning is processed, activating hyper attention (Csernoch & Dani, 2017). The webpage opened in a browser would reveal whether the data are organized in a table-like formation or not (Figure 7). In this sense, the LOL board is quite promising. At first sight, the only distracting factor is the presence or absence of pictures in the second column, and the small icons in the third column. The table can also be suspected of redundancy, since all the records contains the words 'votes', 'Comments', and 'Views'.

At this scanning stage, activating mixed attention (Csernoch & Dani, 2017), we can begin to consider the number of data and their data types stored in the records. The second column is not clear, since it holds several pieces of text-data: title, username, server (in parenthesis), source, and time(Figure 8). The question is how these data are stored and separated. The first column is also tricky, because, by pointing on the votes, the numbers of positive and negative votes are presented instead of the total votes (Figure 8). Apart from these distracting factors and concerns, the webtable seems promising, and convertible.

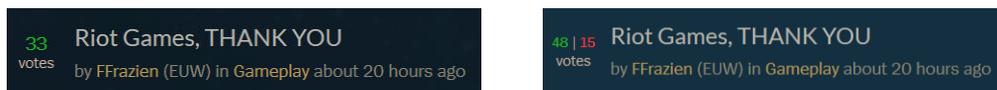

*Figure 8. The change in the way in which the number of votesis presented in the LOL Board.*

One further concern of the conversion process is the selection of applications. The LOL board webtable selected for the present paper, can be converted both in word processors (Section 4) and spreadsheet applications (Section 5), using the same data analysing techniques and algorithms, but requiring different background knowledge, tools, and skills to carry out the conversion.

### 4. KNOWLEDGE-TRANSFER ITEMS – WORD PROCESSOR (WP)

### 4.1. Saving and opening the webpage (WP)

*Table 1. Saving and opening the webpage in a browser and in a word processor, respectively.*

|        | Algorithm | Tool |
|--------|-----------|------|
| 4.1.1. | opening and saving webpage in a browser | Save Page As |
| 4.1.2. | converting the webpage to a word document | starting MS Word→File→Open→File→Save as→File type: Word Document |
| 4.1.3. | deleting the contents outside of a table | selection of contents above and below table→deleting content |

After opening the webpage in a word processor and deleting the contents above and below the table, a relatively clean text table is obtained(Figure 9).This form of the table reveals a couple of unexpected features. The votes – positive, negative, and total –are stored in different paragraphs; however, this can easily be handled by conversion. The column with the icons, the comment category, reveals no problem in this form of presentation, since all the data-cells are filled with text, substituting the icons or their absence, and most importantly, not one cell is left empty. Furthermore, as was suspected, we are faced with redundancy; however, this can be handled rapidly and easily. Finally, disturbing Space characters– primarily data closing spaces – must be handled in order to obtain pure data.

*File types (extensions):* Web Page complete (html), Word Document (docx)

*Knowledge-transfer items:* opening a datafile in a non-assigned application (starting an application, opening a datafile, Ctrl+O, optional but fast), selecting and deleting contents, key combinations for selection (Ctrl+Shift+Home and Ctrl+Shift+End), file conversion: webpage→word document (Save As).

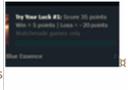

*Figure 9. The LOL board webtable is opened in a word processor.*

However, the second (with picture, the third) column is not arranged according to the data. The title is in a separate paragraph, in accordance with our expectations. The username and the server share another paragraph, but they can be separated along the parentheses. The source and the time share one further paragraph. However, this case is more complicated than it seems, since in this table there are four different ways to indicate the time: "about an hour ago", "a day ago", "about … hours ago", "… days ago" (not presented in Figure 9 but handled in Formula 27–Formula 34). Finally, for unknown reasons, without checking the page source and/or info– i.e. by applying deep attention (Csernoch & Dani, 2017) –, we are surprised by an extra paragraph for the word "by".

**4.2. Arranging/re-arranging columns (WP)– mixed attention mode**

The second (with picture, the third) column seem frightening due to the presence or lack of a picture. In the case an uncertain number of columns it is worth moving these columns to the far right (Figure 10).The new order of columns is the following: (A–1) category, (B–2) new comments, (C–3) views, (D–4) votes, and (E–5 / 5½) title, for short.

*Knowledge-transfer items:* recognizing the number of columns, recognizing a column as an object, selecting columns, moving columns

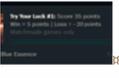

*Figure 10. The order of the columns is changed in the word processor to move the problematic columns to the furthestright position.*

In the current phase of the conversion it is worth knowing the number of records. To count them, we can insert an extra column and set up automated numbering. After registering the record-number we can undo these last steps. The downloaded LOL Boards consist of 966 records (Figure 11).

*Knowledge-transfer items:* records, number of records, inserting a new column, changing column-width, selecting and deleting a column, automated numbering, undo, and stack

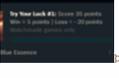

*Figure 11. Counting the numbers of records with automated numbering.*

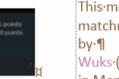

*Figure 12. Automated bullets are removed from votes.*

The final action in this phase does not change the number of columns, but it is better to handle this problem here, specifically by removing the automated bullets (automated numbering with symbols).

*Knowledge-transfer items:* selecting a column, turning off automated numbering/bullets (one of the most important knowledge-transfer items is the similarities between automated numbering and bullets)

### 4.3. Separating data (WP)– deep attention mode

In the following, without giving long explanations, we provide the algorithm, the tools, and the file types to carry out the real data conversion. Separating the algorithm from the tools brings various advantages, as is obvious in programming. The advantages of building algorithms are the following. Algorithms are interface independent, can be constructed and written in any natural language, and inscribed on any surface; tools, on the other hand, are application specific. Since one of our purposes is to support knowledge-transfer, separating the algorithms from the tools plays a crucial role. With this approach various coding solutions can be carried out at various stages of development, and in various applications and versions, without rebuilding the algorithm.

*Table 2. The algorithm and the tools to build up the records of the datatable by removing varying layout solutions of the webtable, presented in Figure 12 (the conversion to a text file and back to a word document is not detailed). Results presented in Figure 13.*

|        | Algorithm                                             | Tool                                                                                                                     |
|--------|-------------------------------------------------------|--------------------------------------------------------------------------------------------------------------------------|
| 4.3.1. | changing paragraphs to data fields                    | replace: `paragraph`→`tabulator`                                                                                         |
| 4.3.2. | deleting all formats and pictures                     | converting the word document to a text file and then back to a word document                                             |
|        | changing column separators                            | Table Tools→Layout→Convert to Text→Separator: Tabs                                                                       |
| 4.3.3. | deleting duplicate tabulator characters               | replace: `tabulatortabulator`→`tabulator`                                                                                |
| 4.3.4. | deleting data-closing space characters                | replace: `spacetabulator`→`tabulator`                                                                                    |
|        |                                                       | replace: `spaceparagraph`→`paragraph`                                                                                    |
| 4.3.5. | changing column separators(back to table) (Figure 14) | Select all →Insert→Table→Convert Text to Table→10 columns and 966 rows (recognized by Word and correct)                  |

*File types (extensions):* Word Document (docx), text file (txt)

*Knowledge-transfer items:* replace, copy, escape sequences for end of paragraph and tabulator characters, saving the file with a new name, deleting formats and pictures by saving as a text file, different separator characters of tables, refreshing files (MS Word and Excel do not do this automatically), Ctrl+F4, Ctrl+O (both are optional but fast), file conversion with Save as, a range of commands

```
Rioter·Comments → 51·new·Comments → 6.4k·Views→56·|·2   →    54 →  votes→Oh·
boy,·time·to·open·a·Hextech·chest→by →  Kanzler·(NA)   →   in·Memes·&·Games·
about·20·hours·ago¶
Boards·Developer·Comments  →   20·new·Comments → 8.5k·Views→72·|·2   →    70
   →   votes→This·mission·pretty·much·sums·up·ranked·matchmaking·by →  Wuks·
(NA)→in·Memes·&·Games·about·23·hours·ago¶
```

*Figure 13. Data-fields are separated with the Tabulator character where data-closing Space characters are deleted.*

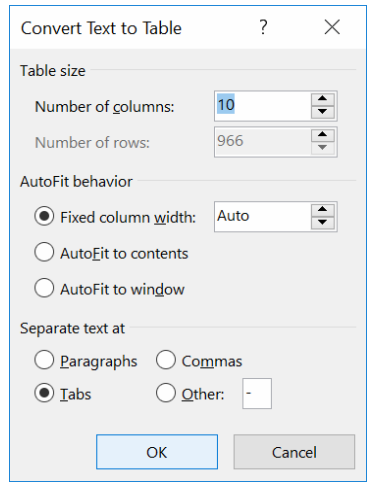

*Figure 14. 10 columns and 966 records are recognized by the word processor.*

*Table 3. Separating positive and negative votes.*

|  | Algorithm | Tool |
|---|---|---|
| 4.3.6. | separating positive and negative votes | replace: `space\|space→tabulator` |

The separation of positive and negative votes evokes several knowledge-transfer items. This is the first action where the number of records plays a crucial role. If the replacement is carried out in the text form, the number of replacements does not match the number of records (Figure 15), indicating that there are other instances of the searched string, apart from the votes. Consequently, this command can only be carried out when the column is selected (Figure 16).

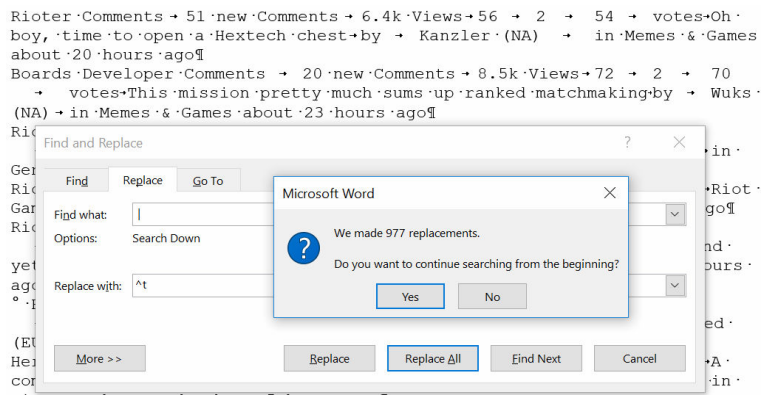

*Figure 15. The number of replacements does not equal the number of records when the replacement is carried out in the whole document.*

One further concern is whether or not to continue the search in the rest of the document. By default, the "Yes" option is offered (Figure 16). In most cases, students do not read the message, they do not pay attention to it, and accept what is suggested by the "almighty" Microsoft. This is one of the knowledge-transfer items which must be taken seriously throughout computer problem-solving: no one else but the user, who is working on the problem, can decide how to continue. Regardless of the interface, students must be trained that all messages must be read, and action must be carried out according to the algorithm previously set up.

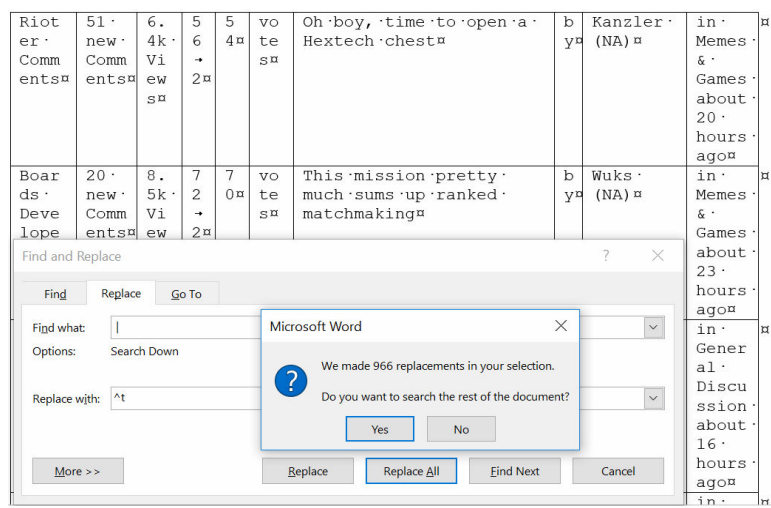

*Figure 16. The number of replacements is equal to the number of records. The replacement is restricted to the selection, so further searching is not allowed.*

*Knowledge-transfer items:* replace, escape sequence of the Tabulator character, number of replacements, changing separator tools in a table, selecting a column, deciding to search the rest of the document

### 4.4. Handling redundancy (WP)– deep attention mode

Redundancy can also be handled with replacements in both word processors and spreadsheets. In the present section the word-version is described.

*Table 4. The algorithm and the tools to reduce redundancy in the table. The numbers in parentheses indicate the numbers of replacements carried out.*

|        | Algorithm | Tool |
|--------|-----------|------|
| 4.4.1. | deleting columns *votes* and *by* | selecting columns→deleting columns |
| 4.4.2. | deleting *in*space | replace: inspace→nothing (966) |
| 4.4.3. | deleting space*Views*(Column C) | replace: spaceViews→nothing (966) |
| 4.4.4. | deleting space*New Comments*(Column B) | replace: spaceNew Comments→nothing (802) |
| 4.4.5. | deleting space*New Comment*(Column B) | replace: spaceNew Comment→nothing (164) (802 + 164 = 966) |
| 4.4.6. | deleting space*Comments*(Column A) | replace: spaceComments→nothing (965) |
| 4.4.7. | deleting space*Comment*(Column A) | replace: spaceComment→nothing (1) (965 + 1 = 966) |

One further aspect of reducing redundancy and separating data with replacements is to build up schemata of Find and/or Replace by applying various contexts. With this method drilling is carried out, with a high number of repeating opportunities. Beyond offering practising environments for Find/Replace, these activities require creativity and, not infrequently, a strict execution order of commands, skills which are extremely important in computer problem-solving.

*Knowledge-transfer items:* replace, selecting and deleting a column, range of commands, order of commands

## 5. KNOWLEDGE-TRANSFER ITEMS – SPREADSHEET APPLICATION (SP)

### 5.1. Saving and opening the webpage (SP)–mixed attention mode

As mentioned above, building up the datatable requires the same algorithm in both applications, described above, with some minor changes.

*Table 5. The algorithm for cleaning the webtable in Excel (conversion to a text file and back to a word document is not detailed). Results presented in Figure 17.*

|        | Algorithm | Tool |
|--------|-----------|------|
| 5.1.1. | opening and saving the webpage in a browser | Save Page As |
| 5.1.2. | converting the webpage to a spreadsheet document | starting MS Excel→File→Open→File→Save as→File type: Excel Workbook |
| 5.1.3. | deleting contents outside of the table | selection of rows above and below the table, deleting rows (right click→local menu→ Delete or Home→Delete→Delete Sheet Rows) |
| 5.1.4. | deleting all formats and pictures | converting the excel document to a text file |
| 5.1.5. | converting the text file to an excel document | Save as: File type: →Excel Workbook |

The algorithm and the tools for cleaning the webtable do not differ at all in the two applications (Table 1 vs. Table 5). The only tool which should be taken into consideration is that in spreadsheets above and below the table it is not only contents, but objects (rows) which must be deleted in the first step of the cleaning process (Table 5).

*File types (extensions):* Web Page, complete (html), Excel Workbook (xlsx), text file (txt)

*Knowledge-transfer items:* saving and converting files with Save As, opening a file manually in a non-assigned application, selecting rows, deleting rows, refreshing a text file manually

### 5.2. Creating records (SP)– deep attention mode

After the webtable is cleaned of formats and pictures, the rearranging of data columns and creating of records must be dealt with. The table consists of Column A for the votes (positive, negative, and total), Columns B and C for the title, user, server, source and time (Columns B and C: originally without and with picture, respectively) (Figure 7), Column D for category, Column E for the number of new Comments, and Column F for the number of Views (Figure 17).

*Figure 17. The cleaned webtable opened in Excel.*

In the cleaned table, all the records occupy four rows (Figure 17, Worksheet LOL). To create one single record from the four rows is a modulo problem. Knowing the record number (3864/4=966; 3864=number of data-rows in Worksheet LOL) allows us to calculate the original position of the data by multiplying the new record number by 4. Within the range of a foursome the specified rows can be reached by a subtraction (Figure 18: Worksheet LOL_rec, Table 6: 5.2.2).

Similar to the solution in word processors (Section 4.3, Step 4.3.1), the next step in the conversion process is building up the records (Table 6). The output is presented in Figure 18. After generating the record numbers, a search in a vector must be completed to write out the data in the corresponding fields. Furthermore, to create the title, user_server, and the source_time columns a condition must be set up to handle the two title columns, due to the optional pictures.

|   | A | B | C | D | E | F | G | H | I |
|---|---|---|---|---|---|---|---|---|---|
| 1 | Rrecord | Rvotes_pn | Rvotes | Rtitle | Ruserserver | Rsourcetime | Rcategory | Rnewcomments | Rviews |
| 2 | 1 | 56 \| 2 | 54 | Oh boy, time to ope| Kanzler (NA) | in Memes & Games about 20 | Rioter Comments | 51 new Comments | 6.4k Views |
| 3 | 2 | 72 \| 2 | 70 | This mission pretty r | Wuks (NA) | in Memes & Games about 23 | Boards Developer Comments | 20 new Comments | 8.5k Views |
| 4 | 3 | 31 \| 2 | 29 | Yeah, we still need b | BeatzBoyFTW (NA) | in General Discussion about 1 | Rioter Comments | 24 new Comments | 4.9k Views |
| 5 | 4 | 48 \| 15 | 33 | Riot Games, THANK | FFrazien (EUW) | in Gameplay about 20 hours a | Rioter Comments | 102 new Comments | 7.6k Views |
| 6 | 5 | 48 \| 17 | 31 | According to op.gg , | ValiantKiller (NA) | in Gameplay about 21 hours a | Rioter Comments | 93 new Comments | 7.4k Views |
| 7 | 6 | 41 \| 0 | 41 | [Lunar Revel] Missio | Matixed (EUNE) | in Report a Bug about 24 hour | Rioter Comments | 43 new Comments | 10.6k Views |
| 8 | 7 | 11 \| 2 | 9 | A comment with me | OtakuBurrito (NA) | in Discuss the Boards about 5 | Herald Comments | 27 new Comments | 1.7k Views |
| 9 | 8 | 63 \| 12 | 51 | I would rather have | BigFBear (EUW) | in Gameplay a day ago | Rioter Comments | 69 new Comments | 9.3k Views |
| 10 | 9 | 18 \| 0 | 18 | Master Yi, a popular | affably evil (NA) | in Report a Bug about 13 hour | Rioter Comments | 12 new Comments | 3.5k Views |

*Figure 18. Data arranged into records with Excel functions.*

The comparison of the two solutions reveals that while the tools applied in word processors are based on general operating system knowledge-transfer items and some specifications frequently used in programming, the spreadsheet-solution requires items brought from mathematics and the ability to handle Lookup & Reference functions.

*Table 6. The algorithm for creating the records in Excel. The fieldnames starting with LOL or R are in sheets LOL or LOL_rec, respectively (Figure 17 and Figure 18).*

|        | Algorithm | Tool |
|--------|-----------|------|
| 5.2.1. | Numbering records (Rrecord) | Formula 1 |
| 5.2.2. | positive and negative votes (Rvotes_pn) | Formula 2 |
| 5.2.3. | total votes (Rvotes) | Algorithm 5.2.2 |
| 5.2.4. | title (Rtitle) | Formula 3 |
| 5.2.5. | user and server (Ruserserver) | Algorithm 5.2.4 |
| 5.2.6. | source and time (Rsourcetime) | Algorithm 5.2.4 |
| 5.2.7. | category (Rcategory) | Algorithm 5.2.2 |
| 5.2.8. | #new comments (Rnewcomments) | Algorithm 5.2.2 |
| 5.2.9. | #views (Rviews) | Algorithm 5.2.2 |

*Knowledge-transfer items:* counting rows and records, modulo, searching in a vector, writing out vector items, setting up a condition, output values depending on the condition, composite functions, array formulas (solutions presented in Table 6)

{=ROW(Rrecord)-1}
*Formula 1: Section 5.2.1*

{=INDEX(LOLvote, Rrecord*4-3)}
*Formula 2: Section5.2.2*

{=IF(INDEX(LOLtitle1,Rrecord*4-3)=0,
   INDEX(LOLtitle2,Rrecord*4-3),
   INDEX(LOLtitle1,Rrecord*4-3))}

*Formula 3: Section 5.2.4*

### 5.3. Separating data (SP)– deep attention mode

One further aspect of building up the records is the separation of data. In the spreadsheet, the negative and positive votes are separated with functions (Table 7, Steps 5.3.1 and 5.3.2), using the same separator character as the word processors (Table 3). In a similar way, based on the same algorithm, the separation of the users and servers are carried out with composite functions (Table 7, Steps 5.3.3 and 5.3.4).The replacement process is not detailed in the word processor for the reason that the two problems require similar, but not identical, algorithms. Applying these algorithms both supports building schemata (called forth later on with fast thinking) and concept-based problem solving.

*Table 7. The algorithm for creating records and data fields using Excel functions (Figure 18andFigure 19).*

|        | Algorithm                                          | Tool      |
|--------|----------------------------------------------------|-----------|
| 5.3.1. | separating the positive votes–left to \| character | Formula 4 |
| 5.3.2. | separating negative votes–right to \| character    | Formula 5 |
| 5.3.3. | separating the user – left to ( character          | Formula 6 |
| 5.3.4. | separating the server –left to ) character         | Formula 7 |

*Figure 19. Data arranged into records and data fields with Excel functions.*

{=LEFT(Rvotes_pn, SEARCH(" | ", Rvotes_pn)-1)*1}
*Formula 4: Section 5.3.1*

{=RIGHT(Rvotes_pn, LEN(Rvotes_pn)-SEARCH(" | ", Rvotes_pn)-1)*1}
*Formula 5: Section 5.3.2*

{=LEFT(Ruserserver, SEARCH("(", Ruserserver)-2)}
*Formula 6: Section 5.3.3*

{=LEFT(RIGHT(Ruserserver,  LEN(Ruserserver)-SEARCH("(",  Ruserserver)),
   LEN(Ruserserver)-SEARCH("(", Ruserserver)-1)}
*Formula 7: Section 5.3.4*

*Knowledge-transfer items:* recognizing separator characters, building up algorithms, handling Text functions, composite functions, discussion, debugging

### 5.4. Handling redundancy (SP)– deep attention mode

As presented in Section 4.4, handling redundancy is conveniently achieved with a sequence of replacements. In spreadsheets, the replacement of characters can be carried out either using the Replacement command (detailed in Section 4.4) or functions. In this section, to avoid repetition, the latter version is presented applying the SUBSTITUTE() function with the empty string as the replacement string. We must note here that there are cases when the TRIM() and/or the CLEAN() functions also work in the cleaning process, but they are not as reliable as the SUBSTITUTE() function. Building functions for reducing redundancy is not complicated. However, the numbers of replacements are not displayed; consequently, checking the number of replacements requires further solutions and skills.

*Table 8. Algorithm for reducing redundancyin categories, using Excel functions (Figure 18 and Figure 19).*

|        | Algorithm | Tool |
|--------|-----------|------|
| 5.4.1. | deleting space*Comments*(Column Ncategory) | Formula 8 |
| 5.4.2. | checking the number of replacements | Formula 9 or Formula 10 or Formula 12 (Figure 20, Cells C2, C3) |
| 5.4.3. | colouring the cells which do not match the requirements (optional) | Formula 13 (Figure 20, Column B) |
| 5.4.4. | deleting space*Comment*(Column Ncategory) | Formula 11 |
| 5.4.5. | checking the number of replacements | Formula 12 (Figure 20, Cells 4) |

To check the number of replacements, we can calculate the number of those records which do not match our requirements. To highlight the non-matching cells conditional formatting can be applied. In the example presented a separate sheet (LOL_test) is set up for the discussion (Figure 20).

Ncategory: {=SUBSTITUTE(Rcategory, " Comments","")}
*Formula 8: Section 5.4.1*

{=SUM(IF(ISERROR(RcategoryT), 1))} (Figure 20, C2)
*Formula 9: Section 5.4.2*

{=SUM(IF(ISERROR(SEARCH(" Comments", Rcategory)),1))} (Figure 20, C3)
*Formula 10: Section 5.4.2*

Ncategory: {=SUBSTITUTE(SUBSTITUTE(Rcategory, " Comments",""),
" Comment","")}
*Formula 11: Section 5.4.4*

{=SUM(IF(ISERROR(SEARCH(" Comment", Ncategory)),1))} (Figure 20, C4)
*Formula 12: Section 5.4.2*

RcategoryT: {=SEARCH(" Comments", Rcategory)} and
conditional formatting: =ISERROR(B2:B976) (Figure 20, Column B)
*Formula 13: Section 5.4.3*

*Figure 20. Testing the number of substitutions with Excel functions and conditional formatting.*

Calculating and colouring in the number of failed replacements reveals the instances which do not match the string of the SUBSTITUTE() / SEARCH() functions, and the mismatching cell(s) can be easily recognized. In this case, similar to the procedure in Word, one failed replacement was found (Comments vs. Comment) (Figure 20, Cells C2 and C3; Sections 4.4.6 and 4.4.7; Table 4).

*Table 9. Algorithm for reducing redundancy in new comments, using Excel functions (Figure 18 and Figure 19).*

| 5.4.6. | deleting space*new Comments*and converting it to a number (Column Nnewcomments) | Formula 14 |
|---|---|---|
| 5.4.7. | checking the number of replacements | Formula 15 or Formula 16 |
| 5.4.8. | colouring the cells which do not match the requirements (optional) | Formula 13 |
| 5.4.9. | deleting space*new Comment* and converting it to a number (Column Nnewcomments2) | Formula 17 |

{=SUBSTITUTE(Rnewcomments, " new Comments","")*1}
*Formula 14: Section 5.4.6*

Checking the number of failed replacements in *new comments* is somewhat simpler than in *categories*, because the conversion from a string datatype to number fails when a substitution is not carried out (Figure 19, Sheet LOL_norm, Column Nnewcomments). The shorter formulas can be applied while using additional cells and vectors (Formula 15), without them longer formulas are needed (Formula 16) (Figure 20, Cells E2 and E3).

The test-algorithm of the number of replacements in the case of " new Comments" is faster than in the previous problem. The SUBSTITUTE() function is the same; however, one further step in the conversion from text to number is carried out, by a multiplication. The mathematical operator returns an error massage if non-convertible text is found (Figure 20).

NnewcommentsT: {=Nnewcomments} and {=SUM(IF(ISERROR(NnewcommentsT), 1))}
*Formula 15: Section 5.4.7*

{=SUM(IF(ISERROR(SEARCH(" new Comments",Rnewcomments)), 1))}
*Formula 16: Section 5.4.7*

Proceedings of the EuSpRIG 2019 Conference "Spreadsheet Risk Management" ISBN : 978-1-905404-56-8
Copyright © 2019, EuSpRIG European Spreadsheet Risks Interest Group (www.eusprig.org) & the Author(s)
Page 16/22

```
{=IF(ISERROR(SEARCH(" new Comments", Rnewcomments)),
    SUBSTITUTE(Rnewcomments, " new Comment","")*1,
    SUBSTITUTE(Rnewcomments, " new Comments","")*1)}
```
*Formula 17: Section 5.4.9*

*Table 10. Algorithm for reducing redundancy in views, using Excel functions (Figure 18 and Figure 19).*

| 5.4.10. | deleting spaceViews (Column Nviews) | Formula 18 |
|---|---|---|
| 5.4.11. | checking the number of replacements | Formula 19 or Formula 20 or Formula 21 |
| 5.4.12. | colouring the cells which do not match the requirements (optional) | Formula 13 |
| 5.4.13. | converting text to number | Table 11 |

{=SUBSTITUTE(Rviews, " Views","")}
*Formula 18: Section5.4.10*

{=SUM(IF(ISERROR(RviewsT), 1))}
*Formula 19: Section 5.4.2*

{=SUM(IF(ISERROR(SEARCH(" views", Rviews)),1))}
*Formula 20: Section 5.4.2*

{=SUM(IF(ISERROR(SEARCH(" views", Nviews)),1))}
*Formula 21: Section 5.4.2*

*Table 11. Algorithm for converting Views to number datatype, using Excel.*

| 5.4.14. | searching for *k* string | Formula 22 |
| 5.4.15. | checking the absence of *k* string | Formula 23 |
| 5.4.16. | separating views based on the absence or presence of *k* string and removing *k* string | Formula 24 |
| 5.4.17. | converting string data type to number datatype | Formula 25 |

Views are presented both as whole numbers (Figure 11, Records 1 and 2; Figure 19) and strings – a real number followed by *k* (kilo), marking a thousand (Figure 11, Record 966; Figure 19). To convert all the numbers of views into a number datatype, we must search for the *k* string, check its presence, remove it, convert the string datatype to a number datatype, and multiply the small numbers by 1 and the *k* numbers by 1000.

{=SEARCH("k", Nviews)}
*Formula 22: Section5.4.10*

{=ISERROR(SEARCH("k", Nviews))}
*Formula 23: Section5.4.10*

{=IF(ISERROR(SEARCH("k", Nviews)), Nviews, SUBSTITUTE(Nviews, "k", ""))}
*Formula 24: Section5.4.10*

{=IF(ISERROR(SEARCH("k", Nviews)), Nviews*1, SUBSTITUTE(Nviews, "k", "") *1000)}
*Formula 25: Section5.4.10*

**5.5. Handling leading space characters– deep attention mode**

After deleting the redundant expressions, the leading Space characters must be deleted. The leading Space characters are recognizable in Excel and Word (Figure 18 and Figure 19, Column I; Figure 21 Column A; Formula 26). The WP-conversion (Figure 21) clearly reveals that one of the Space characters is a non-breaking Space, which must be taken into consideration while calling the SUBSTITUTE() function. Copying the original string into argument position is one solution, while another is the calling of the CHAR() function with 160 as the argument. However, this latter solution requires further knowledge-transfer items.

*Table 12. Deleting leading Space characters in categories, using Excel.*

| 5.5.1. | deleting leading Space characters | Formula 26 |
| | copying both non-breaking and normal Space characters into argument position | 2nd argument of SUBSTITUTE() function |

| | 1 | 2 | 3 | 4 | 5 | 6 | 7 | 8 | 9 |
|---|---|---|---|---|---|---|---|---|---|
| | | | | | | | | ago¤ | |
| Rioter¤ | 102¤ | 7.6k¤ | 48¤ | 15¤ | 33¤ | Riot Games, THANK YOU¤ | FFrazien (EUW)¤ | Gameplay about 20 hours ago¤ | ¤ |
| Rioter¤ | 93¤ | 7.4k¤ | 48¤ | 17¤ | 31¤ | According to op.gg , Karth sitting at 55% Winrate in 3 lanes and yet no nerfs¤ | ValiantKiller (NA)¤ | Gameplay about 21 hours ago¤ | ¤ |
| ° Rioter¤ | 43¤ | 10.6k¤ | 41¤ | 0¤ | 41¤ | [Lunar Revel] Missions won't complete (imgur.com)¤ | Matixed (EUNE)¤ | Report a Bug about 24 hours ago¤ | ¤ |
| Herald¤ | 27¤ | 1.7k¤ | 11¤ | 2¤ | 9¤ | A comment with me quoting Rhaast was removed¤ | OtakuBurrito (NA)¤ | Discuss the | ¤ |
| | | | | | | | | ago¤ | |
| ° Rioter¤ | 144¤ | 38.4k¤ | 36¤ | 2¤ | 34¤ | New Roadmap!So again after one year,Reav3 can you give us another new update on the tier 1 VGU list? (youtube.com)¤ | Lyaso (EUW)¤ | Story, Art, & Sound 4 days ago¤ | ¤ |
| Rioter¤ | 0¤ | 649¤ | 1¤ | 0¤ | 1¤ | Tag de club ne marche pas¤ | XADCX5 (EUW)¤ | Report a Bug 2 days ago¤ | ¤ |

*Figure 21. Leading space characters in categories (Column A).*

Ncategory: {=SUBSTITUTE(SUBSTITUTE(SUBSTITUTE(Rcategory, " Comments", ""),"  Comment", "")," ", "")}
*Formula 26: Section 5.5.1*

*Knowledge-transfer items:* recognizing leading Space characters, differentiating normal and non-breaking Space characters, copying string of Space characters, replacement command or SUBSTITUTE() function

### 5.6. Separating source and time– deep attention mode

There are four different forms of presenting time in the table: (1) about an hour ago, (2) a day ago, (3), about * or ** hours ago (one or two digit-long), and (4)* days ago (only one-digit-long) (Section 4.1).

All these forms must be checked and handled separately to separate the source and the time and to normalize the time, in which process all the presented time-forms are converted to numbers, using the hour as the general unit.

Without going into detail, the formulas of the source and time conversions are presented in Formula 27–Formula 34.To shorten the formulas named vectors were introduced, referred to as Nsource1, Ntime1, Nsource2, Ntime2, Nsource3, Ntime3, Nsource4, and Ntime4.

Nsource1:
{=IF(ISERROR(SEARCH(" about an hour ago", Nsourcetime)),
    Nsourcetime,
    LEFT(Nsourcetime, SEARCH(" about an hour ago", Nsourcetime)-1))}
*Formula 27*

Ntime1:
{=IF(ISERROR(SEARCH(" about an hour ago", Nsourcetime)),
    Nsourcetime,
    1)}
*Formula 28*

Nsource2:
{=IF(ISERROR(SEARCH(" a day ago", Nsource1)),
    Nsource1,
    LEFT(Nsource1, SEARCH(" a day ago", Nsource1)-1))}
*Formula 29*

Ntime2:
{=IF(ISERROR(SEARCH(" a day ago", Ntime1)),
    Ntime1,
    24)}
*Formula 30*

Nsource3:
{=IF(ISERROR(SEARCH(" about", Nsource2)),
    Nsource2,
    LEFT(Nsource2, SEARCH(" about", Nsource2)-1))}
*Formula 31*

Ntime3:
{=IF(ISERROR(SEARCH(" about", Ntime2)),
    Ntime2,
    1*LEFT(RIGHT(Nsource2, LEN(Ntime2)-SEARCH(" about", Ntime2)-
    6),LEN(RIGHT(Nsource2, LEN(Ntime2)-SEARCH(" about", Ntime2)-6))-LEN("
    hours ago")))}
*Formula 32*

Nsource4:
{=IF(ISERROR(SEARCH(" days ago",
    Nsource3)),
    Nsource3, LEFT(Nsource3, SEARCH(" days ago", Nsource3)-3))}
*Formula 33*

Ntime4:
{=IF(ISERROR(SEARCH(" days ago", Ntime3)),
    Ntime3,
    24*LEFT(RIGHT(Ntime3, LEN(Ntime3)-SEARCH(" days ago", Ntime3)+2)))}
*Formula 34*

## 6. CONCLUSION

Present paper describes how word processors and/or spreadsheet applications and/or any combination of these applications can be utilized to convert a webtable to a normalized datatable in educational environment. However, we must emphasise that the subject is highly relevant in commercial and professional organisations too.

In the conversion process, first surface/hyper then mixed or deep attention data analysis is carried out. Following the data analysis, the algorithms of the conversion process are set up. Based on the algorithm, we can decide which tools need to be applied to solve the problems and to carry out the steps of the algorithm. One further consideration of the conversion process is the discussion and debugging phase, when the outputs must be checked and corrected if discrepancies are revealed. All these steps are detailed and discussed in the paper, presenting solutions for the conversion of one of the LOL Boards both in Word and Excel.

In the teaching-learning process presenting authentic tables has a strong motivating effect. However, we must keep in mind that it is the teacher's responsibility which form is introduced in class. The form and content of the table must match the students' background knowledge, stored in long-term memory– originated in informatics and/or other subjects, sciences –, the goals of the classes, and how knowledge gained through these tasks would help them in their further studies. The webtable→datatable conversion processes open up novel methods for developing various computer problem-solving and computational thinking skills by utilizing knowledge-transfer items.

## 7. ACKNOWLEDGEMENT

The project was supported by the European Union, co-financed by the European Social Fund (EFOP-3.6.3-VEKOP-16-2017-00002).

Walkenbach, J. (2010) Excel 2010 Bible. Wiley Publishing, Inc., Indianapolis, IN.

Katz, A. (2010) Beginning Microsoft Excel 2010. Apress.

## 9. SOURCES